# Al$_{0.68}$Sc$_{0.32}$N/SiC based metal-ferroelectric-semiconductor capacitors operating up to 1000 °C

Yunfei He[1], David C. Moore[2], Yubo Wang[1], Spencer Ware[1], Sizhe Ma[1], Dhiren K. Pradhan[1], Zekun Hu[1], Xingyu Du[1], W. Joshua Kennedy[2], Nicholas R. Glavin[2], Roy H. Olsson III[1] & Deep Jariwala[1],*

[1] Department of Electrical and System Engineering, University of Pennsylvania, Philadelphia, PA 19104, USA

[2] Materials and Manufacturing Directorate, Air Force Research Laboratory, Wright-Patterson AFB, Dayton, OH 45433, USA

* Corresponding author: Deep Jariwala, dmj@seas.upenn.edu

**Abstract:**

Ferroelectric (Fe) materials-based devices show great promise for non-volatile memory applications, yet few demonstrate reliable operation at elevated temperatures. In this work, we demonstrate Ni/Al$_{0.68}$Sc$_{0.32}$N/4H-SiC metal-ferroelectric-semiconductor capacitors for high-temperature non-volatile memory applications. Our 30-nm thick ferroelectric Al$_{0.68}$Sc$_{0.32}$N film grown on SiC exhibits stable and robust ferroelectric switching up to 1000°C. The coercive field decreases linearly from −6.4/+11.9 MV cm$^{-1}$ at room temperature to −3.1/+7.8 MV cm$^{-1}$ at 800°C, further reducing to −2.5 MV cm$^{-1}$ at 1000°C. At 600°C, the devices achieve remarkable reliability with ~2000 endurance cycles and over at least 100 hours of retention with negligible polarization loss. At 800°C, the devices retain data for at least 10,000 seconds and exceed 400 write cycles. Our results further highlight the potential for ferroelectric AlScN thin-films particularly when paired with SiC semiconductor substrates for high-temperature non-volatile memory.

Nonvolatile memory (NVM) devices are critical for data storage across applications ranging from consumer electronics to industrial systems, where retaining information without power is essential. Various NVM technologies have been developed to meet these needs, including flash memory,[1] resistive random-access memory (RRAM),[2] magnetic random-access memory (MRAM),[3] phase-change memory (PCM),[4] and ferroelectric random-access memory (FeRAM). Among these competitors, ferroelectric-based memory devices have demonstrated superior characteristics, including fast programming time, low writing energy, and excellent endurance.[5]

Despite the growing need for NVM in high temperature applications, thermal stability remains a major obstacle for most NVM technologies, with commercially available products struggling to function reliably at temperatures exceeding 300°C.[6] While widely studied ferroelectric materials like hafnium zirconium oxide ($Hf_{1-x}Zr_xO$) show significant degradation such as — over 80% reduction in remanent polarization at 400°C[7], the wurtzite-structure III-nitride-based ferroelectric material aluminum scandium nitride ($Al_{1-x}Sc_xN$) is known for its large remnant polarization has emerged as a promising alternative [8-12] for NVM applications.[13,14] With its high Curie temperature, AlScN has been reported to retain ferroelectricity after annealing at 1000°C.[15] Furthermore, in our former work, we have already demonstrated the high-temperature operation of the Ni/AlScN(45 nm)/Pt metal-insulator-metal (MIM) structure ferroelectric diode (FeD) NVM devices, showing long retention and high endurance at 600°C.[16]

High-temperature digital circuits are crucial for aerospace, automotive, and energy exploration industries, where components must operate reliably in extreme thermal environments.[17-20] Silicon carbide (SiC) is the model semiconductor material for high temperature electronics due to its high thermal stability and wide-band gap.[21] Thus, SiC-based devices have demonstrated functionality at temperatures up to 800°C, while emphasizing the demands of a memory technology that is directly integrable with SiC.[17,22-26] Building on the successful demonstration of Al/AlScN/SiC-based ferroelectric capacitors in our previous work,[27] we have now designed our devices for enhanced high-temperature performance. This study presents temperature-dependent measurements of Ni/$Al_{0.68}Sc_{0.32}$N/SiC metal-ferroelectric-semiconductor (MFeS) structure ferroelectric capacitors, specifically designed for compatibility with SiC-based integrated circuits, addressing the need for high-temperature digital circuits with embedded memory in extreme thermal environments.

A schematic of the Ni/$Al_{0.68}Sc_{0.32}$N/SiC MFeS capacitor is shown in Figure 1(a). The fabrication process began with sputtering and patterning 200 nm-thick Ni onto a heavily n-type doped 4H-SiC wafer as bottom electrodes, followed by an annealing process to form Ohmic contacts between the Ni electrodes and SiC wafer. A 30 nm $Al_{0.68}Sc_{0.32}$N ferroelectric layer was then co-sputtered directly onto the 4H-SiC wafer via methods previously reported.[15,26] Subsequently, 200 nm-thick Ni top electrodes were sputtered using identical settings to the bottom electrodes and patterned by photolithography into circular pads with a diameter of ~

60 μm. Atomic force microscopy (AFM) analysis of the 30-nm-thick $Al_{0.68}Sc_{0.32}N$ ferroelectric layer on SiC, shown in Figure 1(b), revealed a low surface roughness of 0.74 nm, indicating high-quality growth directly on the 4H-SiC wafer. Figure 1(c) presents a scanning electron microscopic (SEM) image of the patterned top Ni electrode pad, confirming an actual diameter of ~63 μm [See Supplementary Information S1 for detailed SEM image].

To characterize the coercive field and remanent polarization of the fabricated $Ni/Al_{0.68}Sc_{0.32}N/SiC$ ferroelectric capacitors, we employed two methods: bipolar fast sweeping voltage for J-E hysteresis loop measurements and ultrafast pulsing using positive-up negative-down (PUND) measurement. Throughout this work, all measurements maintained a consistent configuration, applying voltage at the bottom electrodes while sensing the current response at the top electrodes. Figure 1(d) shows the J-E hysteresis loop obtained at room temperature under a bipolar triangular voltage sequence with a frequency of 10 kHz. An asymmetric voltage sequence was applied to compensate for the voltage drop across the depletion region of the heavily n-type doped SiC substrate, with significantly higher positive voltage compared to the negative direction.[22] Note that with a higher voltage applied in the positive direction, the leakage current component, due to increased electron injection through the ferroelectric layer, is higher than that in the negative direction [See Supplementary Information S2 for symmetric voltage applied to the device at room temperature, showing high leakage current component in both positive and negative directions]. The J-E hysteresis loop reveal current density peaks indicating ferroelectric switching under both positive and negative applied voltages. The coercive field ($E_C$) was estimated at -6.4 MV cm$^{-1}$ and +11.9 MV cm$^{-1}$ at 10 kHz, consistent with our previous work. [11,14,22] However, substantial leakage current under positive voltage prevented accurate determination of the current density peaks from the ferroelectric contribution and reliable estimation of the remanent polarization (Pr). Therefore, we implemented PUND measurements to determine the remanent polarization.

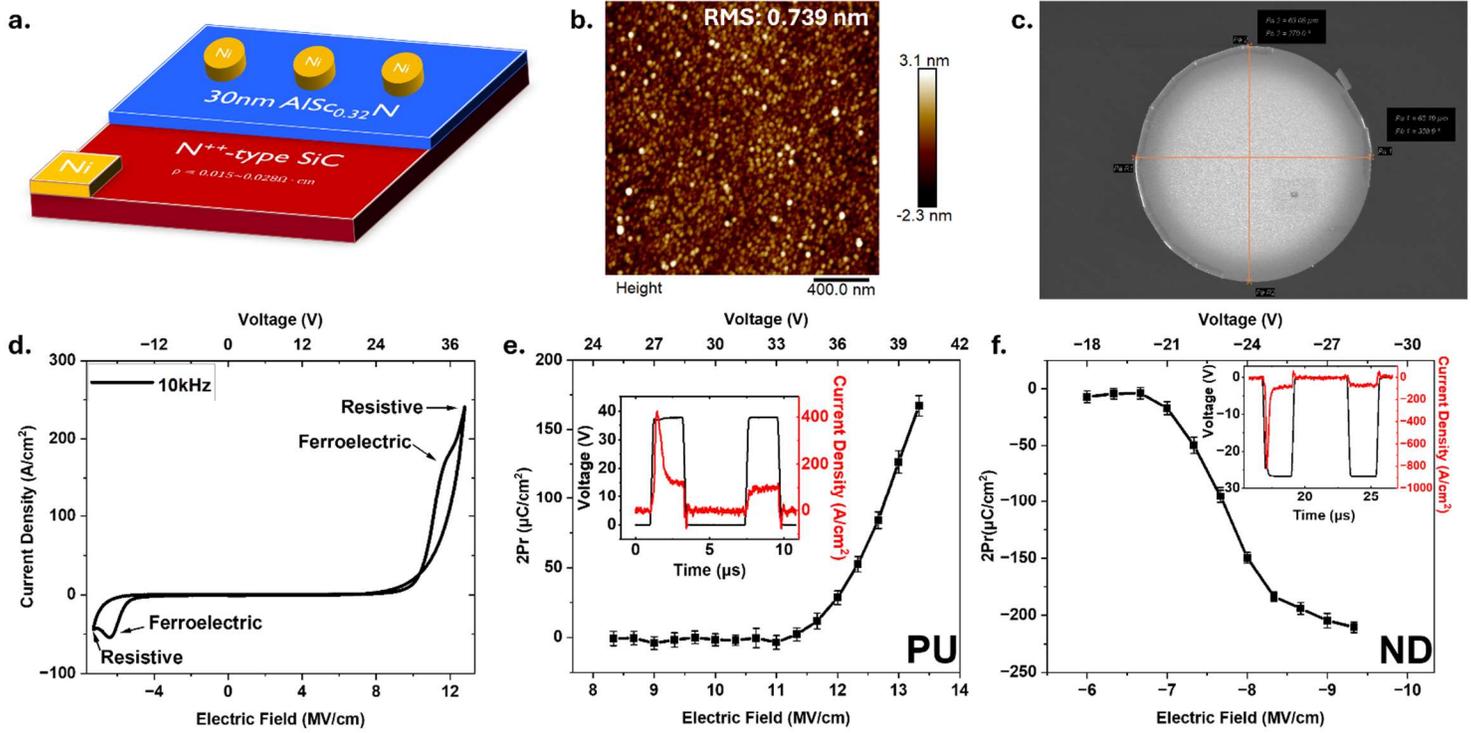

**Fig. 1. (a)** Schematic of the Ni/Al$_{0.68}$Sc$_{0.32}$N (30nm)/4H-SiC metal-ferroelectric-semiconductor (MFeS) capacitor. Both top and bottom electrodes are sputtered Ni with a thickness of 200 nm. **(b)** AFM of the surface topology of the 30-nm-thick Al$_{0.68}$Sc$_{0.32}$N ferroelectric film directly growth on the 4H-SiC wafer, exhibiting a low surface roughness of 0.74 nm. **(c)** SEM image of the top Ni electrode pad of the fabricated device, with an actual diameter of 63.1 μm. **(d)** J-E hysteresis curve of the demonstrated MFeS capacitor at 10 kHz, the black arrows indicate the current density peaks due to leakage current and ferroelectric switching current components, respectively. **(e)** and **(f)** The remanent polarization (2P$_r$) calculated from PUND measurement on ten (10) Al$_{0.68}$Sc$_{0.32}$N/SiC ferroelectric capacitors under different voltage applied in both **(e)** negative and **(f)** positive directions. The 2P$_r$ are normalized by the area of the top electrodes, and the error bar indicates the device-to-device variation of the fabricated capacitors. The inset displays the current response to the PU pulse sequence and the ND pulse sequence, respectively. The current density peaks indicate ferroelectric switching.

The remanent polarization (2Pr) calculated from PUND measurements across ten (10) Al$_{0.68}$Sc$_{0.32}$N/SiC ferroelectric capacitors at varying pulse voltages at room temperature are shown in Figure 1(e) and (f). The insets show transient current responses with corresponding voltage pulses for "PU" and "ND" sequences. For all pulse sequences ("P", "U", "N", and "D"), we employed a rise/fall time of 100 ns and pulse width of 2 μs. Polarization values were obtained by integrating the current response over time for each sequence, followed by subtracting the "U" and "D" pulse sequence polarizations from the "P" and "N" sequences to estimate 2Pr in both voltage directions. Results showed a saturated Pr of −105.2 ±

2.3 µC cm$^{-2}$ for the "ND" sequences at 28 V, matching our previous findings.[22] For "PU" sequences, the large uncompensated leakage current due to asymmetric voltage application resulted in continuously increasing 2Pr with positive voltage. We further developed a modified measurement approach using a triangular voltage sequence with two positive and two negative voltage segments. By subtracting the second positive/negative segment's current response from the first, we obtained clearer current density curves relative to applied electric field, revealing more distinct ferroelectric-related current density peaks in both voltage directions [Also see Supplementary Information S2 for current density characteristics of the demonstrated devices with leakage subtraction].

We investigated the high-temperature performance of the fabricated Al$_{0.68}$Sc$_{0.32}$N/SiC ferroelectric capacitors through J-E hysteresis and PUND (positive-up negative-down) measurements across temperatures ranging from room temperature to 1000°C. All experiments were conducted under vacuum conditions at pressures below 6 mTorr [See Supplementary Information Table S3 for chamber pressure with respect to the temperatures]. Figure 2(a) presents the J-E hysteresis curves measured at 10 kHz for selected temperatures. As temperature increased, we observed a dramatic rise in leakage current, necessitating a reduction in applied voltage to prevent device breakdown at elevated temperatures. Despite this voltage adjustment, ferroelectric switching current density peaks remained observable. At 900°C and above, the leakage current in the positive direction reaches 10 mA, corresponding to the current compliance limit of the measurement tool (indicated by the black dashed line in Figure 2(a)). Notably, in the negative direction, a distinct current density peak attributed to ferroelectric switching is observable at both 900°C and 1000°C. The temperature dependence of the coercive field ($E_C$) is extracted from the J-E curves and is presented in Figure 2(b) calculated by taking the derivative of the current density with respect to the electric field [See Supplementary Information S4-S6 for the detailed calculation]. The $E_C$ exhibits a linear decrease from −6.4/+11.9 MV cm$^{-1}$ at RT to −3.1/+7.8 MV cm$^{-1}$ at 800°C. The $E_C$ further reduces to −2.9 MV cm$^{-1}$ at 900°C and −2.5 MV cm$^{-1}$ at 1000°C, demonstrating a continued reduction in coercive field with increasing temperatures [See Supplementary S7 for breakdown field distribution of five devices at each selected temperature]. Figure 2(c) illustrates the current response from PUND measurements at selected temperatures. To ensure consistency in obtaining saturated remanent polarization at different temperatures, we started from low voltages for the PUND sequences and progressively increased them at each temperature until the calculated remanent polarization reached saturation. This approach allow us to accurately capture the remanent polarization while preventing the increased leakage current from breaking down the devices at high temperatures. [See Supplementary information S8 for PUND voltage sequences].

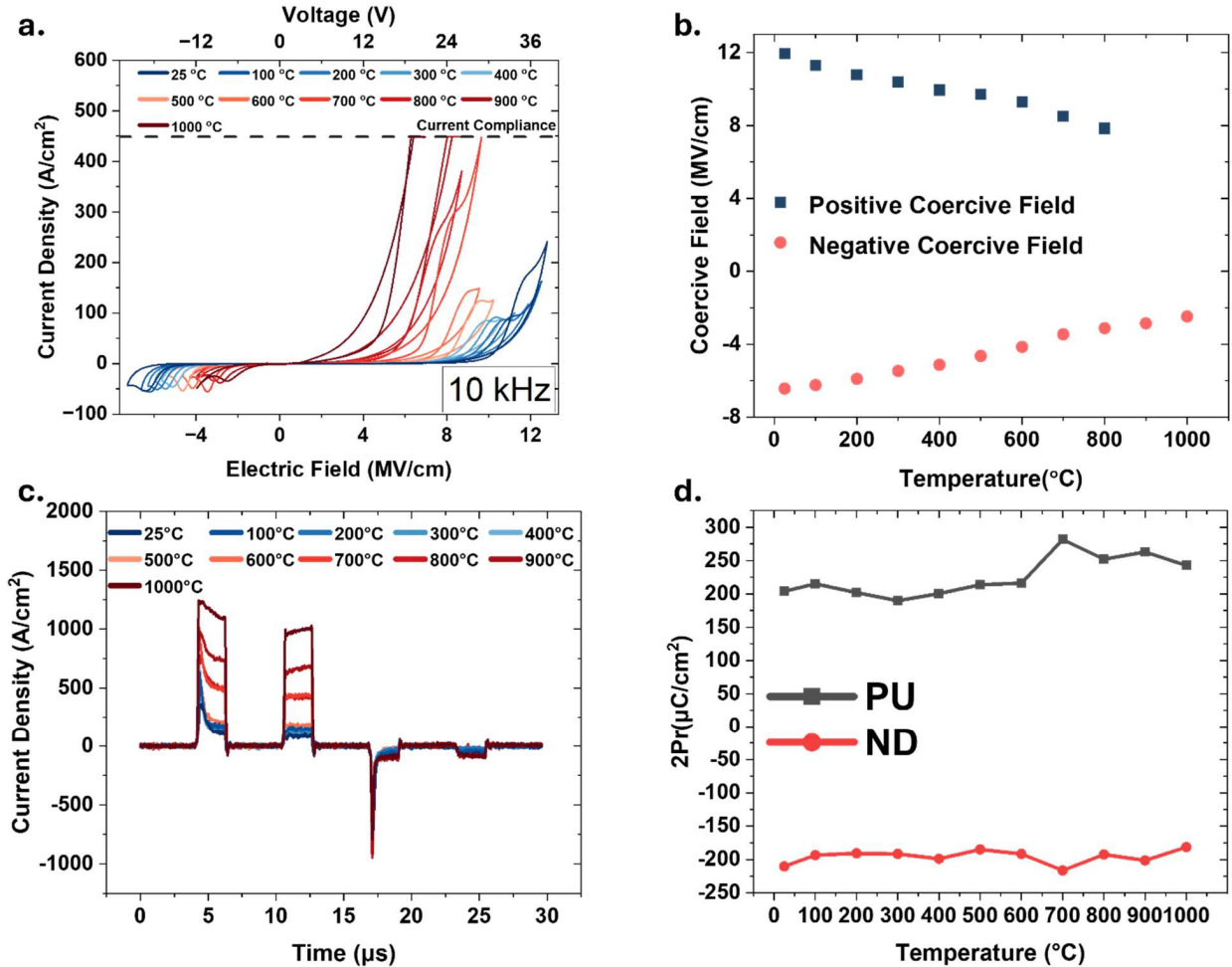

**Fig. 2. (a)** 10 kHz J-E hysteresis loops of the fabricated $Al_{0.68}Sc_{0.32}N$/SiC ferroelectric capacitor at elevated temperatures (T) with a range from RT (25°C) to 1000°C. The applied voltage decreases when increasing the temperature, to accommodate the rising leakage current and to prevent the device from breaking down. The black dashed line indicates the measured current compliance of the tool. **(b)** Coercive field ($E_C$) calculated from the J-E hysteresis loop with respect to the measured temperature. **(c)** Current response of a single asymmetric PUND measurement on the ferroelectric capacitor under elevated temperatures. **(d)** Saturated remanent polarization ($2P_r$) calculated from the current response of the PUND measurements. Note that the $2P_r$ for PU pulse sequence above 600°C increases due to the uncompensated leakage current.

Figure 2(d) presents the saturated remanent polarization ($2P_r$) calculated by integrating the current response from the PUND measurements. The remanent polarization in ND sequences remains relatively stable, with the $P_r$ of −90.6 µC cm$^{-2}$ at 1000°C. On the other hand, the remanent polarization in PU sequences starts rising at 700°C since the uncompensated leakage current increases with temperature in the positive direction. At temperatures of 600°C and lower, the remanent polarization $P_r$ maintains at approximately +108.1 µC cm$^{-2}$, showing consistency with the results obtained from the ND sequences and aligns with our prior results at room temperature.[21]

Through J-E hysteresis curves and PUND measurements, we demonstrate that the $Al_{0.68}Sc_{0.32}N$ film maintains stable ferroelectric characteristics at temperatures up to 1000°C. This thermal stability establishes $Al_{0.68}Sc_{0.32}N$/SiC-based devices as promising candidates for memory applications in high-temperature circuit designs. Given that reliability is crucial for memory devices, particularly in elevated temperature operations, we conducted comprehensive investigations of retention and endurance characteristics at high temperatures, with specific focus on performance at 600°C.

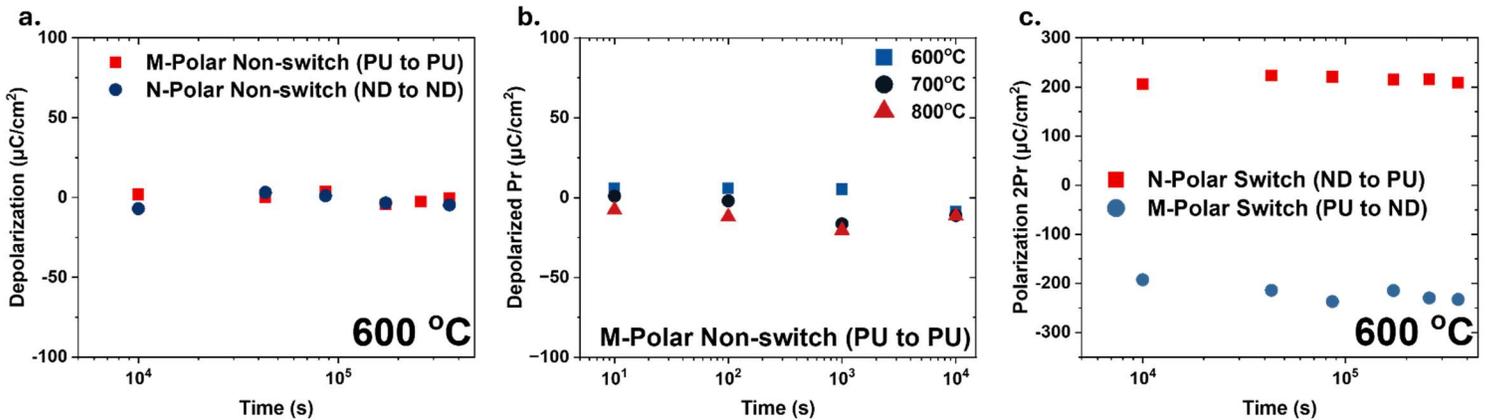

**Fig. 3. (a)** Same state retention performance on the demonstrated $Al_{0.68}Sc_{0.32}N$/SiC-based capacitor in both Metal-Polar (M-Polar) and Nitrogen-Polar (N-Polar) directions at 600°C. **(b)** Same state retention performance in M-Polar direction of the ferroelectric capacitor at selected elevated temperatures. **(c)** Opposite state retention performance in both M-Polar and N-Polar directions at 600°C.

To evaluate the retention performance of the $Al_{0.68}Sc_{0.32}N$/SiC-based capacitors, we conducted both same state and opposite state retention tests based on PUND pulse measurements with different pulse sequence configurations while maintaining the devices at high temperature. [See Supplementary information S9 for details of the non-switching and switching retention pulse sequences.] By integrating the current response from the pulsed voltage sequences, we calculated the amount of depolarized polarization from same state retention measurements and $2P_r$ from switching methods. Figure 3(a) presents an ultra-low depolarization in both Metal-Polar (PU sequence followed by another PU sequence) and Nitrogen-Polar (ND sequence followed by another ND sequence) after $3.6 \times 10^5$ s (100 hours) at 600°C, with a polarization loss of $-0.4/-4.7$ μC cm$^{-1}$ in M-Polar and N-Polar after 100 hours, respectively. Note that the negative depolarization in the M-Polar is mainly due to the leakage current and cycle-to-cycle variation of the demonstrated ferroelectric capacitor. In the N-Polar direction, the polarization depolarized by only 3.4% of its remanent polarization at 600°C, which demonstrates significant retention at the selected temperature. Since the demonstrated $Al_{0.68}Sc_{0.32}N$ ferroelectric thin film was co-sputtered on SiC with an initial polarization direction of N-Polar, it is more susceptible for the ferroelectric film to be depolarized when programmed in the M-

Polar direction. Therefore, we focus our retention analysis on the ferroelectric capacitors in the M-Polar state at higher temperatures. Due to the limitation of our measurement tool, we were not able to maintain the temperature at 900°C and above for a long-time interval. Hence the same state measurement for M-Polar state was measured only up to 800°C. Figure 3(b) shows the same state retention results of the $Al_{0.68}Sc_{0.32}N$/SiC-based capacitors programmed in the M-Polar direction at selected temperatures with a retention time ranging from $10^1$ s to $10^4$ s. Over $10^4$ s, the ferroelectric film retains its polarization in the M-Polar direction with no significant depolarization at the selected temperatures. It is important to note that at 700°C and 800°C, increased leakage current and cycle-to-cycle variation have a greater impact on the calculated depolarization, resulting in further negative depolarization in the M-Polar direction. Since the same state retention measurement in the M-Polar direction yields negative depolarization value, which is inconclusive, we proceeded with opposite state retention measurements on the ferroelectric capacitor at 600°C. These measurements were carried out concurrently on six (6) separate devices, each set with different retention times, with the longest retention time of $3.6 \times 10^5$ s (100 hours), as shown in Figure 3(c). The remanent polarization $2P_r$ remains stable, with values of $-116.2/+104.3$ µC cm$^{-2}$ at $3.6 \times 10^5$ s, compared to $-96.3/+102.8$ µC cm$^{-2}$ at $10^4$ s. This indicates no significant depolarization in either the M-Polar or N-Polar directions, demonstrating excellent retention stability over extended periods at high temperatures.

We continued to evaluate the endurance of the $Al_{0.68}Sc_{0.32}N$/SiC-based ferroelectric capacitor as another aspect of the high temperature reliability of the ferroelectric film. Figure 4(a) presents the endurance cycles based on PUND measurement at selected temperatures, ranging from room temperature (25°C) to 1000°C. To measure the endurance of the fabricated ferroelectric capacitors, we employed a single-positive-single-negative fatigue pulse sequence, with preset iteration cycles between each PUND measurement, where the $2P_r$ was measured. For each fatigue pulse sequence, the same voltage used in the PUND measurement was applied at each temperature. Prior to each endurance measurement at each temperature, multiple PUND measurements with incrementally increasing voltages were conducted, similar to the method described in Supplementary Information 4, to achieve saturated polarization. Once the voltage corresponding to saturated polarization at the selected temperature was determined, it was applied for five PUND pulses as wake-up cycle, then applied for both fatigue and PUND pulses to measure the endurance of the fabricated devices. This approach minimizes the risk of device breakdown from uncompensated leakage current and ensured effective ferroelectric switching. At RT (25°C), the measured device endures $4.9 \times 10^3$ cycles under $-24/+38$ V fatigue pulses with a pulse width of 2 µs, demonstrating improved endurance compared to our previous works. As temperature increases, the cycling endurance decreases, with endurance of 292 cycles under $-12/+29$ V at 900°C, and 183 cycles under $-12/+28$ V at 1000°C. It is important to note that the initial endurance measurement up to 1000°C effectively served as an annealing run for subsequent tests. Following this initial measurement, we conducted a second set of endurance measurements on the same samples, now considered annealed at 1000°C. The annealed devices demonstrate enhanced performance, enduring up to 945 cycles before breakdown at 900 C, indicating robust cycling endurance. [See Supplementary Information S10 for additional endurance results at 900°C and 1000°C of the annealed sample]. We also conducted endurance measurements on nine devices that were maintained at 600°C for over 100 hours, with the applied voltage of −15/+31 V, displayed in Figure 4(b). We conducted the measurements while the devices remained at 600°C, observing a maximum endurance of $2 \times 10^3$ cycles, an average endurance of 1,352 cycles, and a device-to-device variation of 162 cycles. Note that the large contact pads of the measured capacitors cause a notable variation in reliability. Due to the limitations of the high-temperature probe station and probes, we measured only devices with diameters larger than 50 μm, using hard probe tips at elevated temperatures. Our previous works extensively discuss the relationship between ferroelectric capacitor size and device-to-device variation[11,16]. Both retention and endurance tests at high temperatures demonstrate the high reliability of the $Al_{0.68}Sc_{0.32}N$/SiC-based ferroelectric capacitors, positioning AlScN/SiC-based ferroelectric devices as strong contenders for high-temperature memory device applications. Notably, the retention of $10^4$ s and an endurance of 467 cycles at 800°C are sufficient for read-intensive memory applications, such as firmware storage and sensor data logging, which prioritize long retention and reliable operation over high write endurance in elevated temperature environments.

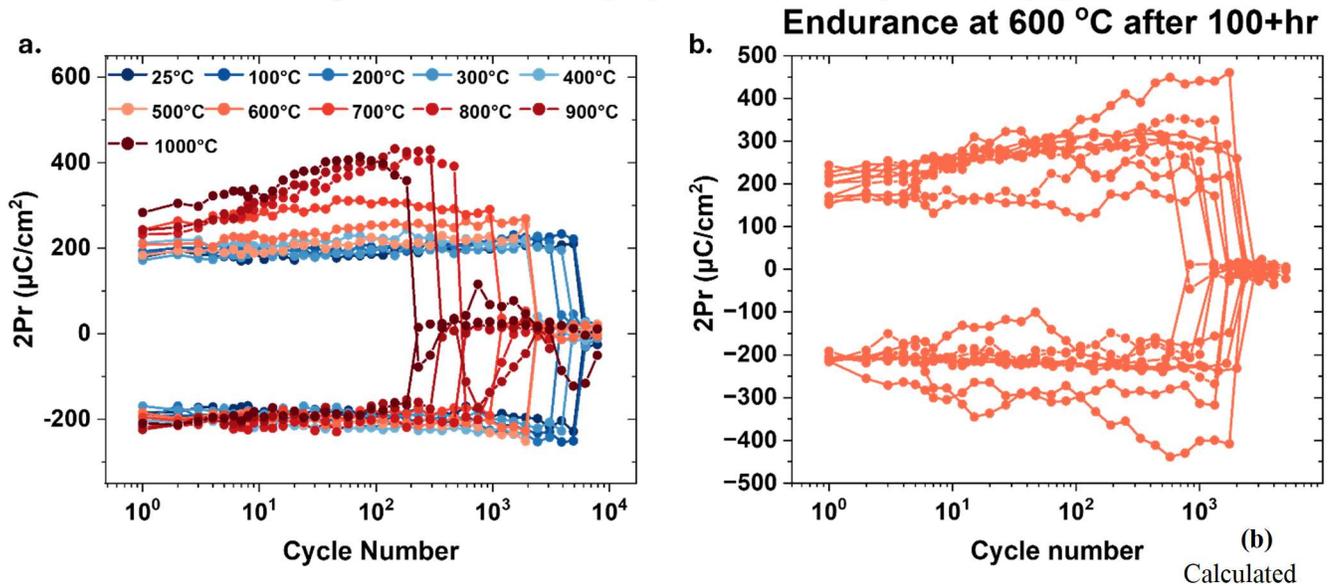

**Fig. 4. (a)** Calculated $2P_r$ of the demonstrated $Al_{0.68}Sc_{0.32}N$/SiC-based capacitors, based on PUND measurement, is shown with respect to the number of fatigue pulses at selected temperatures, ranging from RT to 1000°C. **(b)** Calculated $2P_r$ for nine ferroelectric capacitors at 600°C, with the sample maintained at 600°C for over 100 hours.

In this study, we demonstrate the high-temperature operation of Ni/$Al_{0.68}Sc_{0.32}N$/SiC-based metal ferroelectric semiconductor (MFeS) capacitors. Through J-E hysteresis and PUND measurements, the temperature-dependent properties are evaluated, including coercive field ($E_C$), remanent polarization (Pr), from room temperature up to 800°C, and partial $E_C$ measurements, full PUND measurement as well as cycling endurance measurements extending up to 1000°C. The devices exhibit exceptional high-temperature reliability in both retention and endurance tests. At 600°C, the capacitors maintain stable remanent polarization for $3.6 \times 10^5$ s (100 hours) with negligible depolarization and sustain 2000 cycles of write endurance after 100 hours of heating at 600 °C. Even at 800 °C, the devices demonstrate retention exceeding 10,000 s and write cycling endurance of 467 cycles. These results establish AlScN/SiC-based ferroelectric capacitors as promising candidates for high-temperature non-volatile memory applications. The integration of a 30 nm-thick AlScN ferroelectric thin-film in a MFeS capacitor structure that is robustly switchable and stable at high temperatures establishes a reliable foundation for the future development of AlScN/SiC-based MFeS-FETs, addressing critical needs in high-temperature memory applications and bridging a significant gap in high-temperature integrated circuits for digital computing. Their compatibility with SiC-based logic devices offers a robust and scalable memory solution for extreme thermal environments, paving the way for advancements in reliable, high-temperature digital electronic systems.

## Experimental Section

### $Al_{0.68}Sc_{0.32}N$ Ferroelectric Thin Film Preparation

Thin film of 30 nm $Al_{0.68}Sc_{0.32}N$ was grown directly on the 350 μm thick heavily n-type doped 4H-SiC (Powerway Wafer Co., Limited, Nitrogen N-type Doped, resistivity of $\rho = 0.015\text{–}0.028 \: \Omega \cdot cm$) with a co-sputtering system (Evatec Clusterline 200 II) at 350 °C, with a $N_2$ gas flow of 30 sccm. The Al target and Sc target sputtering powers were set as 1000 W and 655 W, respectively, to create 32% Sc concentration.

### Device Fabrication

Before the ferroelectric thin film was deposited, a 200-nm-thick Ni bottom electrodes was first sputtered (Lesker PVD75 DC/RF Sputterer) and patterned on the corner of the 4H-SiC wafer, following an annealing process at 1050°C for 2 min with Argon flow to form the ohmic contact between the metal and semiconductor interface. After the deposition of the AlScN ferroelectric thin film, 200-nm-thick Ni top electrodes were patterned using a lift-off process and Ni deposition under the same settings described above.

### Device Characterizations

All electrical characteristics measurements, including J-E hysteresis, PUND, retention and endurance tests, were performed by a Keithley 4200A-SCS parameter analyzer, with built-in nonvolatile memory (NVM) library. The tested samples were placed in a vacuum chamber probe station (HP1000V-MPS, Instec Inc), with a heated stage providing selected temperature environments for temperature-dependent measurements. In the J-E hysteresis measurements, we applied a bipolar triangular wave with a frequency of 10 kHz to obtain the temperature-dependent coercive field. For the PUND and reliability measurements, each single pulse in the pulse sequences was a voltage pulse with a rise/fall time of 200 ns and a pulse width of 2 μs, using temperature-dependent voltages to measure and calculate the remanent polarization and reliability of the fabricated ferroelectric capacitors.

## Supplementary Information

Additional experimental details, including detailed SEM image, electrical measurements setup and additional electrical characterization results.

## Acknowledgement

D.J., R.H.O. and Y.H. acknowledge primary support for this work from AFRL via the FAST program D.J. also acknowledges support from the Air Force Office of Scientific Research (AFOSR) GHz-THz program FA9550-23-1-0391. N.R.G. and D.C.M. gratefully acknowledge support from Air Force Office of Scientific Research (AFOSR) GHz-THz program grant number FA9550-24RYCOR011.

## Author Contributions

D.J., R.H.O., Y.H. conceived the concepts of devices, measurements and sample fabrications; Y.H. and Z.H fabricated the samples with assistance from X.D.; the device is measured at RT with assistance from D.K.P., S.M.; Y.H., D.C.M., S.M., Y.W., and S.W. performed all the high temperature measurements; Y.H., D.C.M., and Y.W. analyzed all the electrical data; Y.H. performed AFM measurement; D.C.M. performed SEM measurement; All data analysis are under supervision of D.J., R.H.O.; N.R.G. and W.J.K. guided the project. Y.H. wrote the manuscript. All authors provided their inputs to the paper and Supplementary Information.

# Supplementary Material

# Al$_{0.68}$Sc$_{0.32}$N/SiC based metal-ferroelectric-semiconductor capacitors operating up to 1000 °C

Yunfei He[1], David C. Moore[2], Yubo Wang[1], Spencer Ware[1], Sizhe Ma[1], Dhiren K. Pradhan[1], Zekun Hu[1], Xingyu Du[1], W. Joshua Kennedy[2], Nicholas R. Glavin[2], Roy H. Olsson III[1] & Deep Jariwala[1,*]

[1] Department of Electrical and System Engineering, University of Pennsylvania, Philadelphia, PA 19104, USA

[2] Materials and Manufacturing Directorate, Air Force Research Laboratory, Wright-Patterson AFB, Dayton, OH 45433, USA

[*] Corresponding author: Deep Jariwala, dmj@seas.upenn.edu

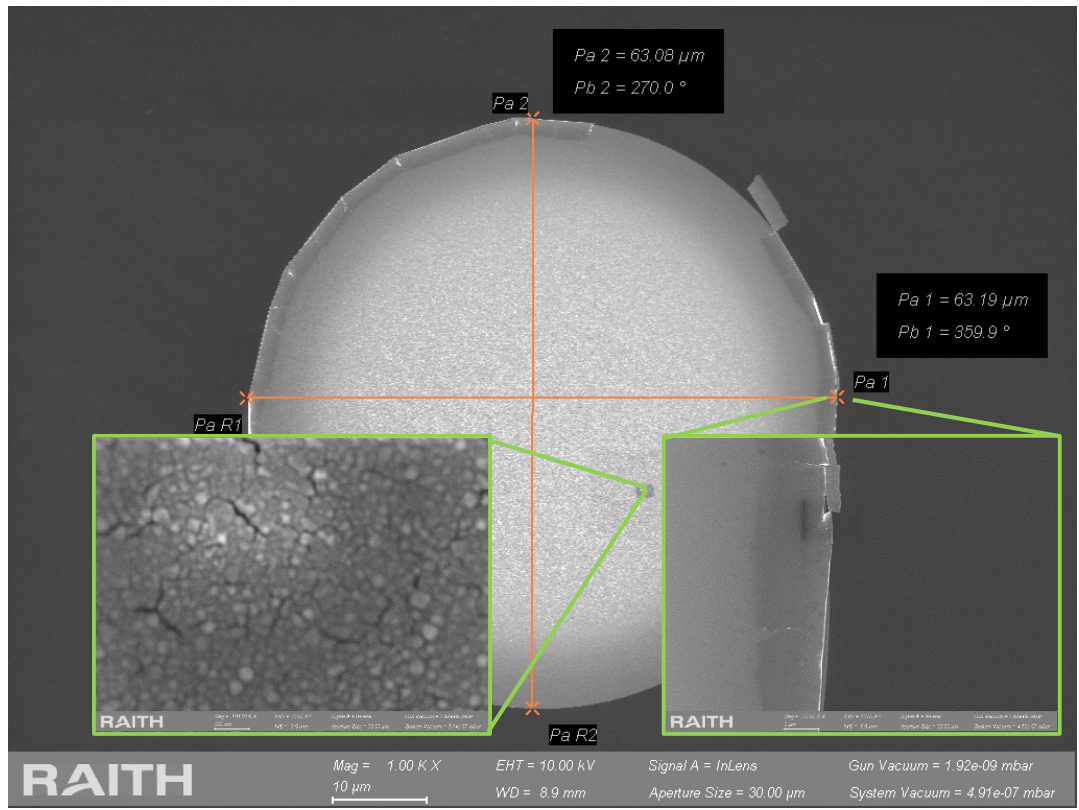

**Figure S1.** Scanning electron microscopy (SEM) image showing the patterned Ni top electrode pad post high T measurements, with the actual diameter of 63.1 μm. The left inset presents a zoomed-in view of the region where the probe was applied. The right inset displays a close-up of the edge of the Ni top electrode, revealing cracks, which are assumed to have been caused by high temperature treatment during the measurements.

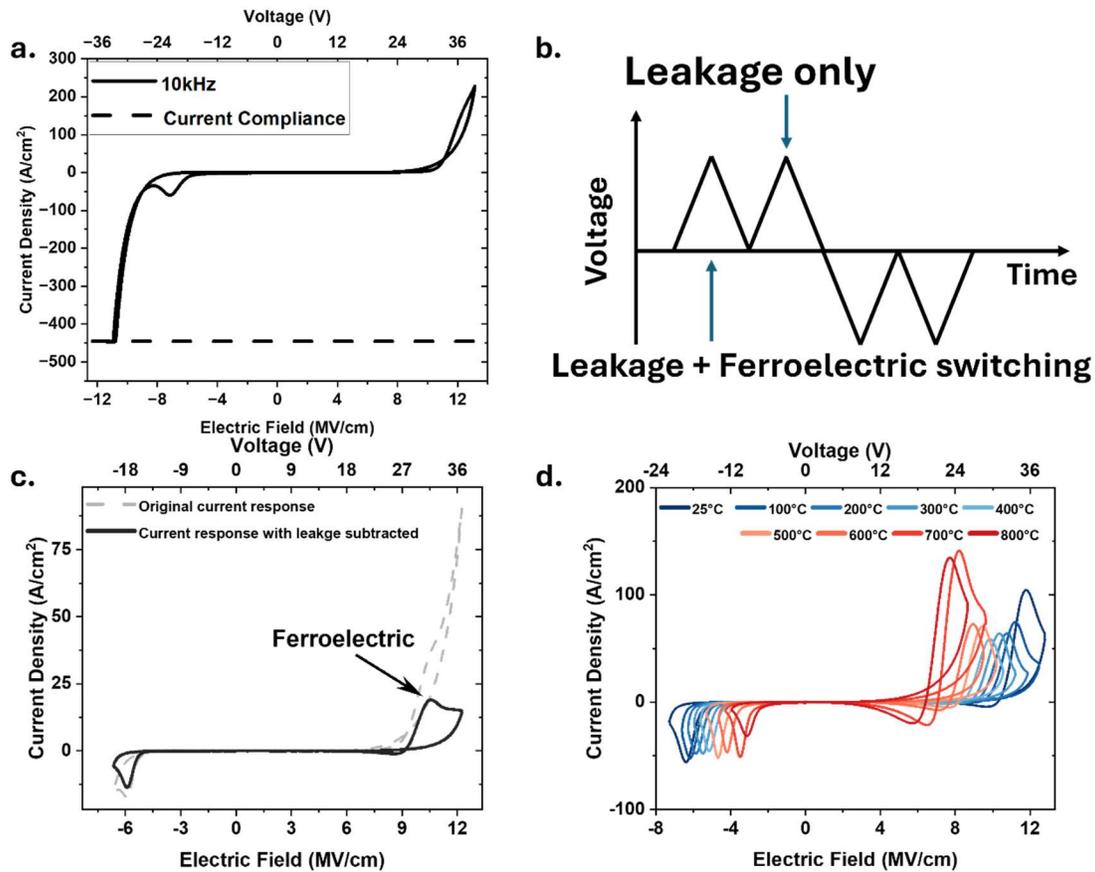

**Figure S2. (a)** The AC J-E hysteresis characteristics of the demonstrated device at room temperature. The black dashed line indicates the measured current compliance of the tool. With a symmetric voltage of −40/+40 V applied, we would observe a high leakage current in both directions. **(b)** A schematic of the wave sequences consists of two positive triangular pulses and two negative triangular pulses, each with a rise/fall time of 25 µs. In the first positive and first negative triangular waves, the current response is expected to include both leakage and ferroelectric switching. In contrast, the second positive and second negative triangular waves are anticipated to show a current response that is primarily due to leakage current. **(c)** a comparison between the original current density response obtained from AC J-E measurement in Figure 2 of the manuscript (dashed black data) and the results with the leakage compensated (solid black data). The leakage compensated results were derived by subtracting the current density response of the second positive/negative triangular waves from the first positive/negative triangular waves. This mathematical approach clarifies the current density peak associated with ferroelectric switching, making it more observable. **(d)** the temperature-dependent current density response was measured at selected temperatures, ranging from RT (25 °C) to 800 °C using the aforementioned leakage compensation method. The current density peaks contributed by ferroelectric switching are clearly observed in both voltage directions at each selected temperature. Note that the negative current density response between 0 MV cm$^{-1}$ and 8 MV cm$^{-1}$ at high temperatures is a result of pure mathematical subtraction. This occurs due to the increased carrier mobility at the Al$_{0.68}$Sc$_{0.32}$N/SiC interface, which enhances the leakage current under the second positive pulse, leading to a negative result from the subtraction.

| Temperature (°C) | Chamber pressure (mTorr) |
|---|---|
| 100 | 3.4 |
| 200 | 3.4 |
| 300 | 3.5 |
| 400 | 3.8 |
| 500 | 4.3 |
| 600 | 4.7 |
| 700 | 5.2 |
| 800 | 5.7 |
| 900 | 5.7 |
| 1000 | 5.8 |

**Table S3.** The temperature-dependent chamber pressure of the probe station shows an increase in pressure as temperature increases. This pressure increase is primarily attributed to the limitations of the vacuum conditions within the probe station chamber.

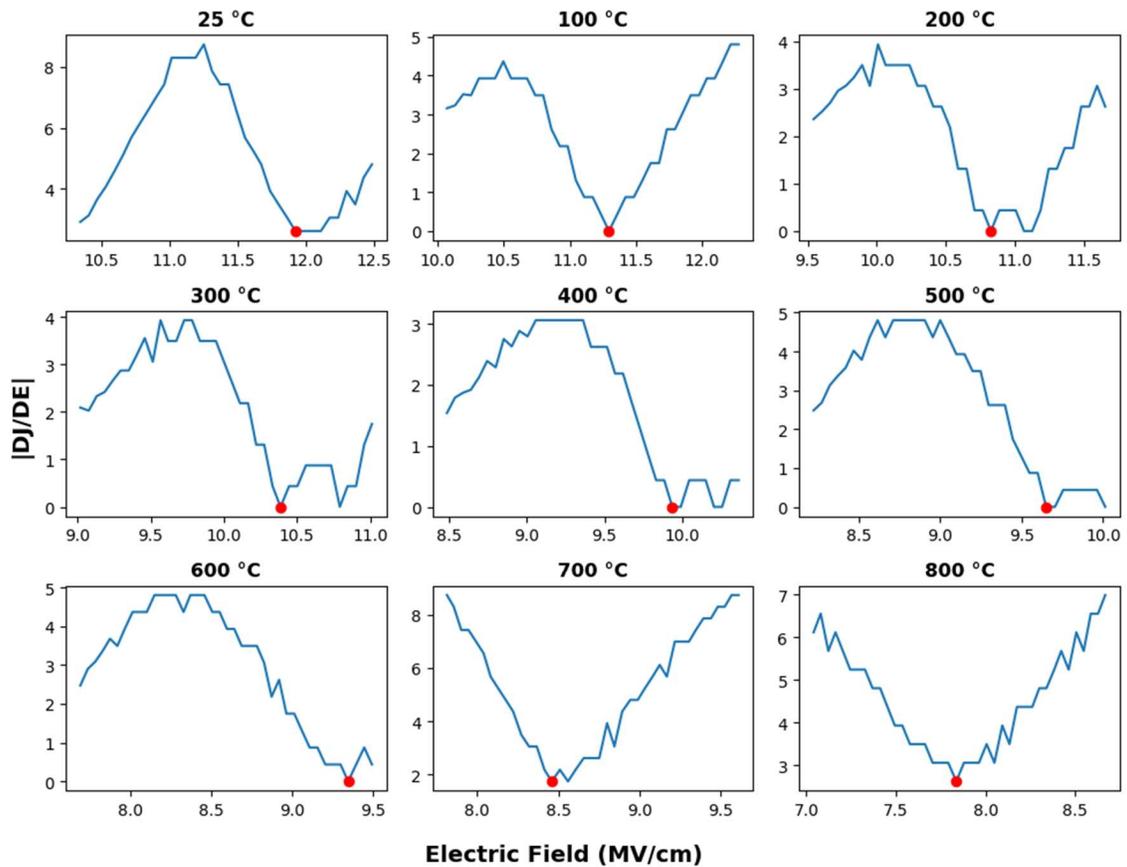

**Figure S4.** Positive coercive field values estimated by locating the minimum absolute values of the derivative of the current density with respect to the electric field from the Figure 2(a) J-E hysteresis loop at different temperatures. The red dot represents the minimum absolute values that indicate the corresponding coercive field.

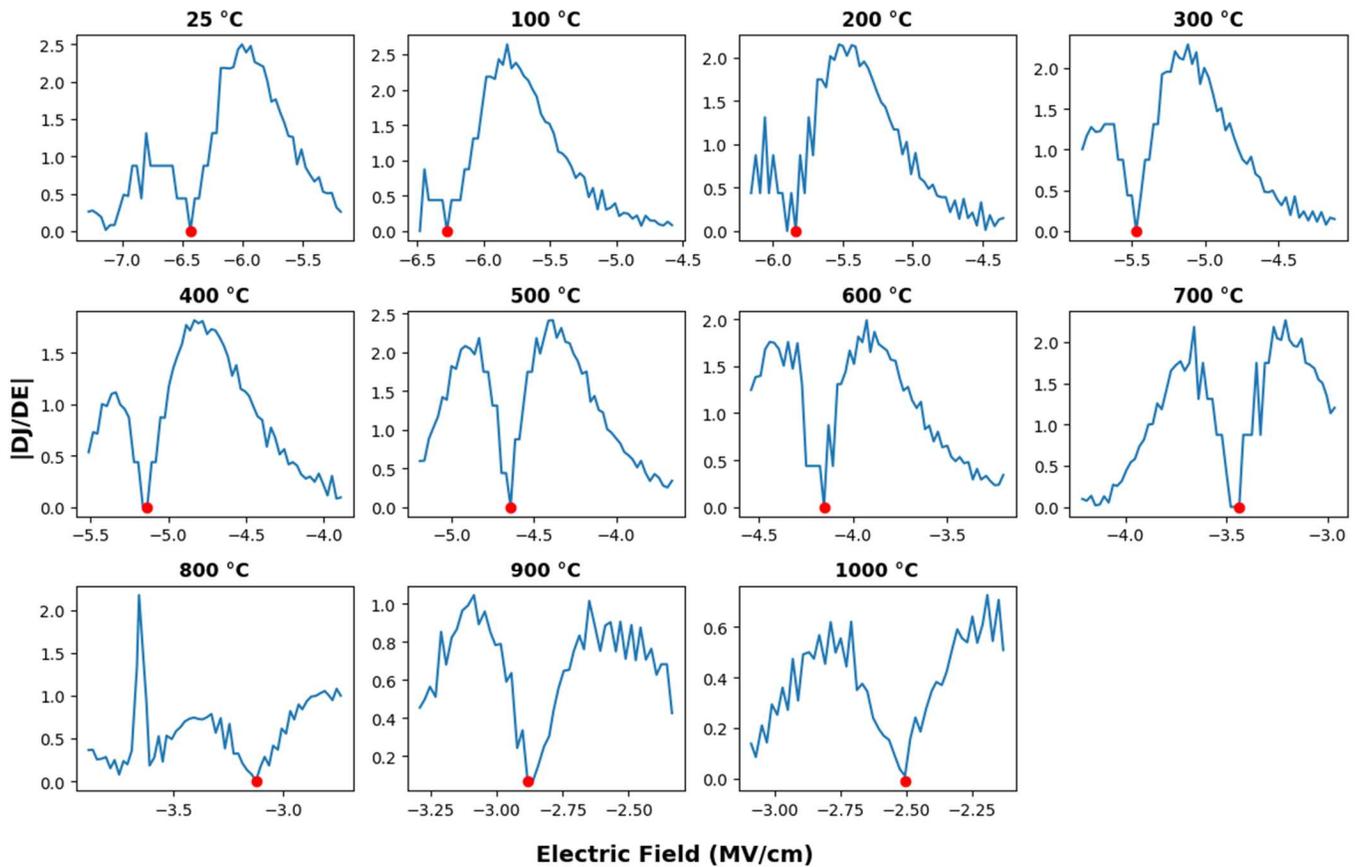

**Figure S5.** Negative coercive field values estimated by locating the minimum absolute values of the derivative of the current density with respect to the electric field from the Figure 2(a) J-E hysteresis loop at different temperatures. The red dot represents the minimum absolute values that indicate the corresponding coercive field.

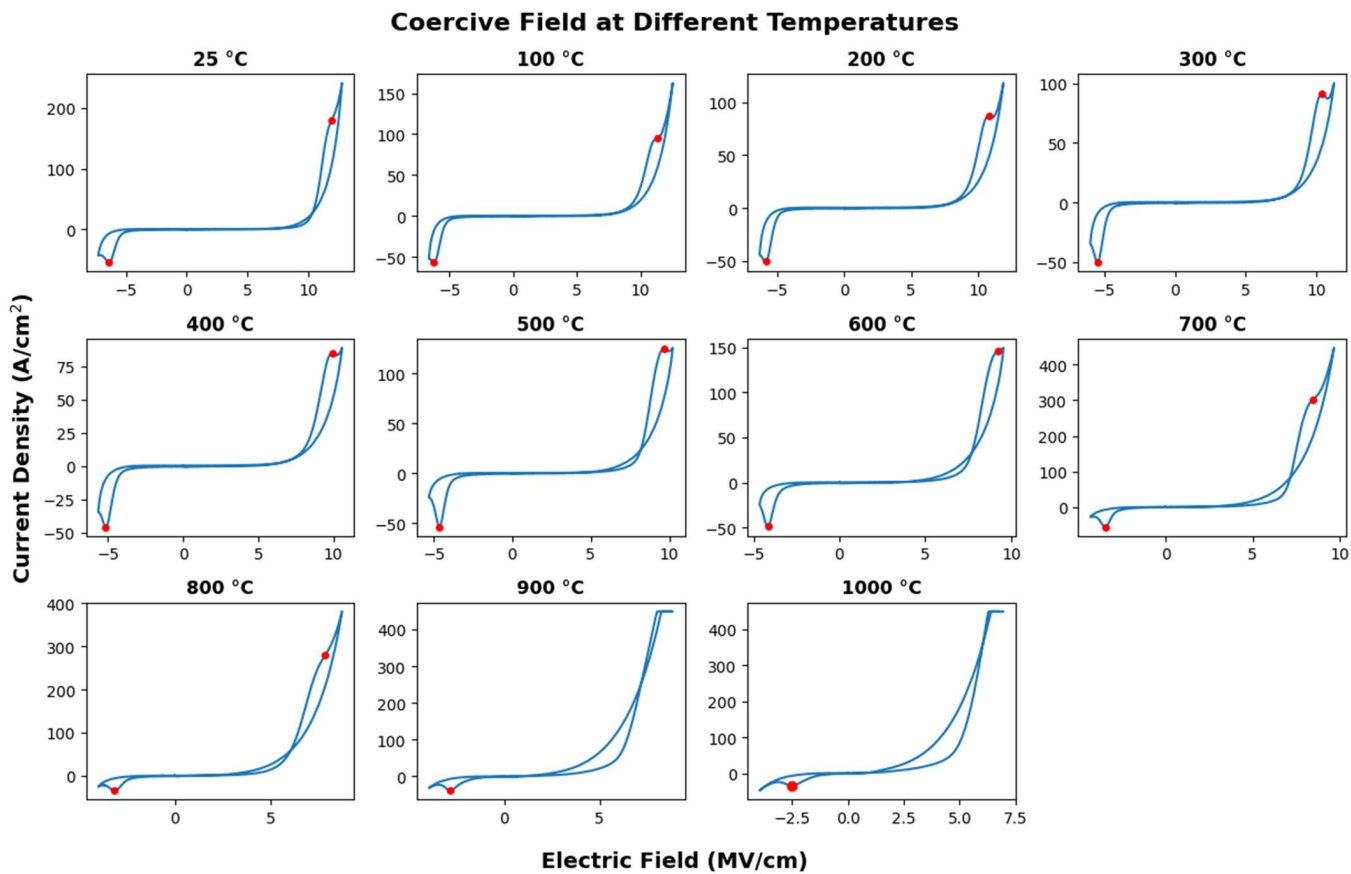

**Figure S6.** Estimated coercive field ($E_C$) values marked by red dots at different temperatures. Note at 900 °C and 1000 °C, the current compliance of 10mA limits the current response at positive direction, where the positive $E_C$ is not marked.

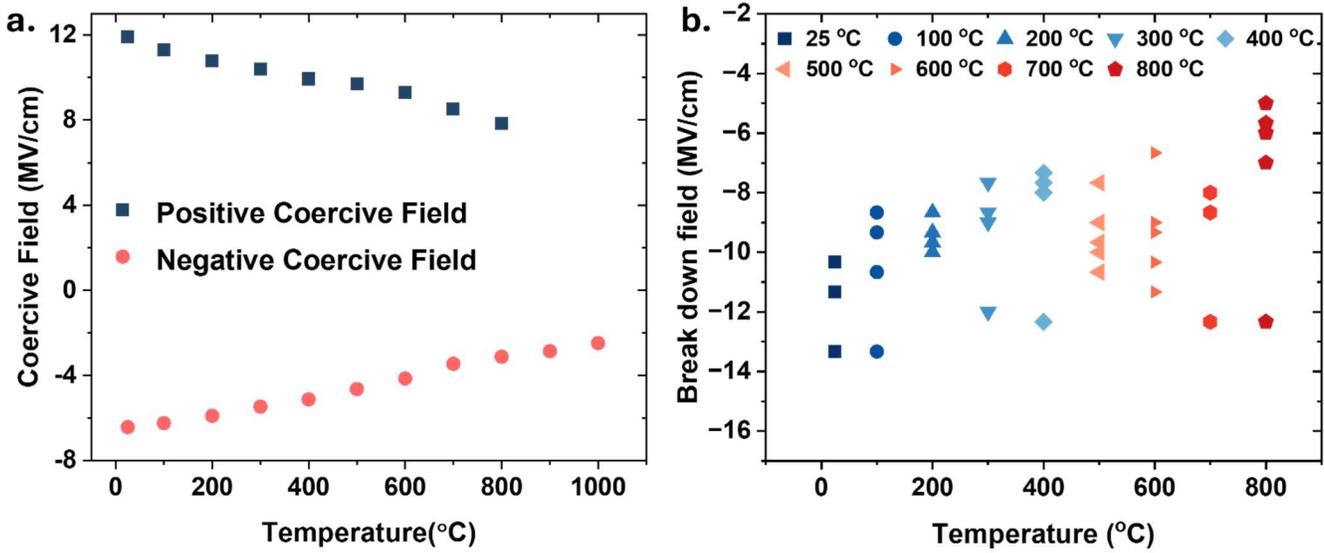

**Figure S7. (a)** Coercive field ($E_C$) calculated from the J-E hysteresis loop with respect to the measured temperature. **(b)** Break down field ($E_{BD}$) of five devices at each selected temperature in the negative voltage directions based on J-E hysteresis loop measurements. To determine the breakdown field during J-E hysteresis measurements, we first apply a voltage sufficient to identify the coercive field in both directions. Subsequently, the voltage is incrementally increased in the negative direction in 1 V steps until the device experiences breakdown, while the voltage applied in the positive direction is maintained same. The electrical field across the device at which breakdown occurs is recorded as the breakdown field. Note that the results show a significant device-to-device variation of the breakdown field. This is due to the large size of the top electrode pads of the demonstrated devices, which has been discussed in detail in our prior works.[1,2]

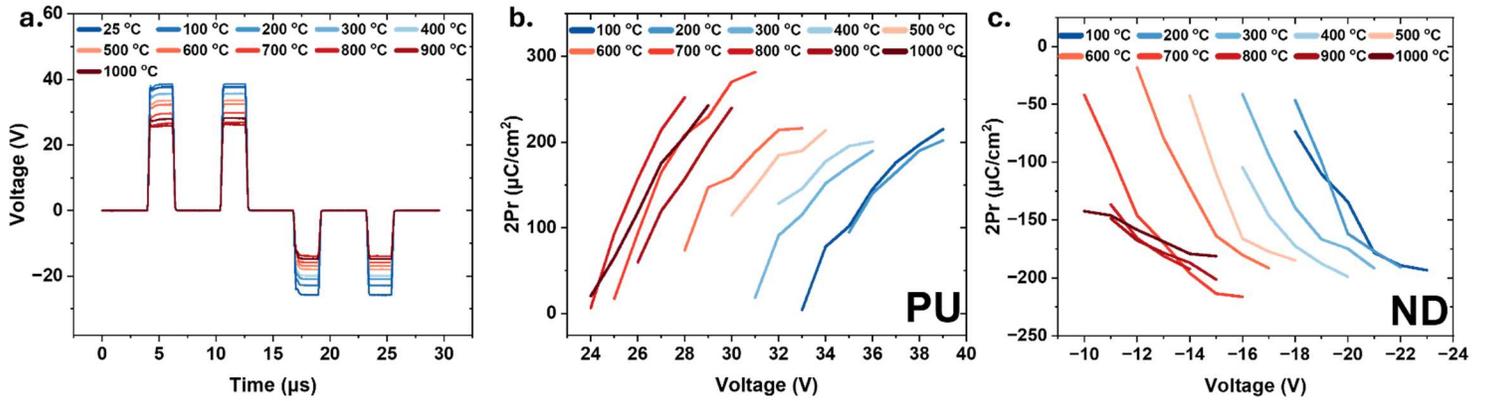

**Figure S8. (a)** The PUND measurement voltage sequences at selected temperatures are presented, corresponding to the current response in Figure 2(c), showing a clear trend of decreasing applied voltage as temperature increases. This adjustment is necessary to prevent the device from breaking down by the thermal-induced leakage current. **(b)** and **(c)** display the remanent polarization ($2P_r$) as a function of the applied voltage in the PUND measurement in **(b)** positive and **(c)** negative directions, respectively. Note that the $2P_r$ in the positive direction (corresponding to PU sequences) increases drastically due to the thermal-induced leakage current. In contrast, a saturation polarization at each selected temperature is clearly observed in the negative direction (ND sequences), indicating stable ferroelectric switching despite the temperature changes.

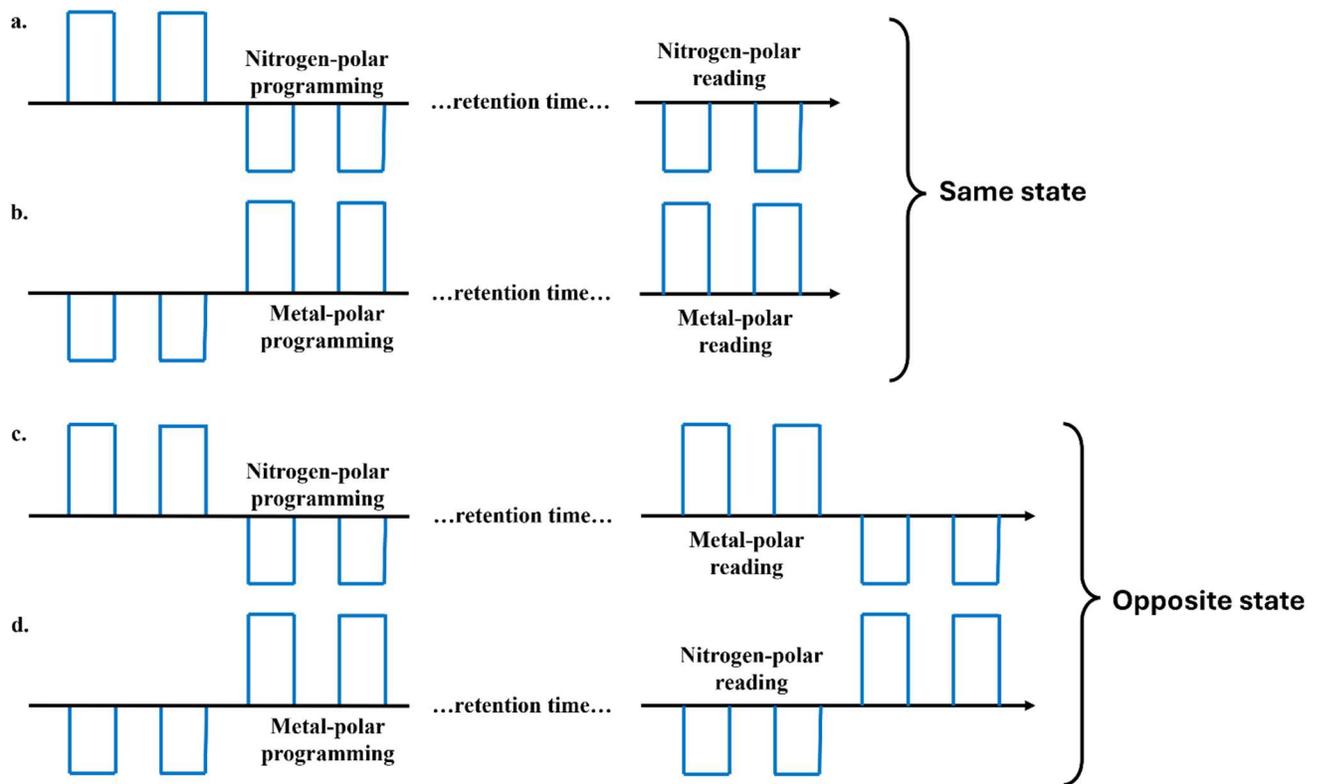

**Figure S9.** The **(a-b)** same state and **(c-d)** opposite state retention measurement pulse sequence schematic using PUND ultra-fast pulsing measurements.[3,4] For same state retention measurements, we first applied a **(a)** PUND (**(b)**NDPU) four pulse sequence to programming the device fully in the Nitrogen-polar (Metal-polar). Then after a certain retention time, we applied a ND (PU) sequence again to obtain the same state non-switched polarization, as the amount of polarization loss. For opposite state retention measurements, we obtained the polarization corresponding to **(c)** N-polar/ **(d)** M-polar after different sets of retention times. Each "P", "U", "N", and "D" pulse sequence applied has a pulse width of 2 µs, rise/fall time of 200 ns, and pulse delay of 4 µs.

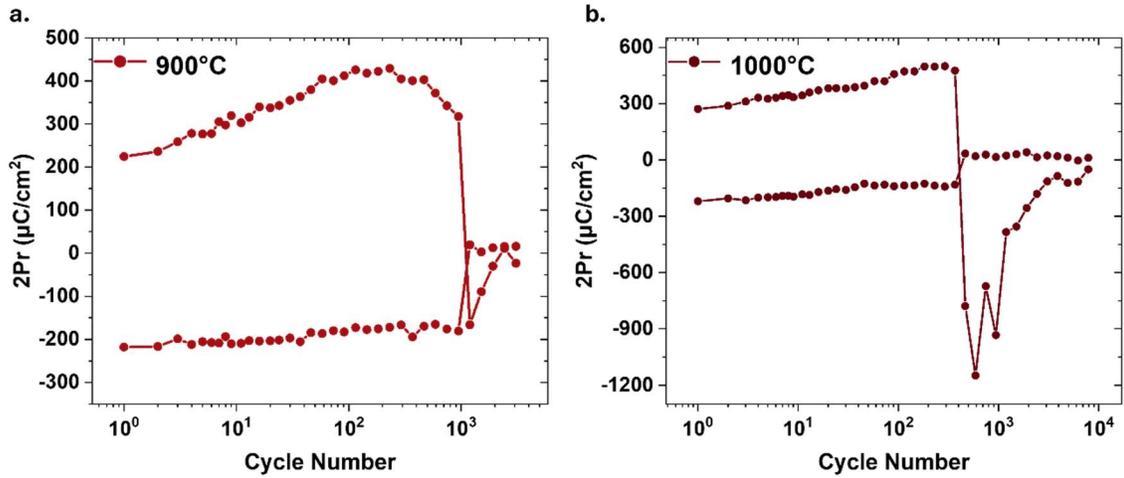

**Figure S10. (a)** Endurance performance of the annealed ferroelectric capacitors at 900°C, demonstrating up to 945 cycles before breakdown following initial annealing at 1000°C. **(b)** Endurance measurement of the same annealed device at 1000°C with an endurance cycle of 365 cycles, highlighting its thermal stability and endurance behavior under repeated high-temperature cycling. Note: when at 1000°C, the device exhibits a significant negative 2Pr value in the PU sequence after breakdown, which is attributed to substantial variations in the current response of the device with a large-area electrode post breakdown.